\documentclass[a4paper,12pt]{mekit09}
\usepackage[T1]{fontenc}
\usepackage[english]{babel}
\usepackage[latin1]{inputenc}
\usepackage[squaren]{SIunits}
\usepackage{amsmath,amssymb}
\usepackage[sf,SF]{subfigure}
\usepackage{tikz}
\usepackage{color}
\usepackage{url}
\usepackage{exscale,times,graphicx}
\usepackage{cleveref}

\crefname{equation}{Equation}{equations}
\creflabelformat{equation}{#2(#1)#3}
\crefname{pluraleq}{Equations}{equations}
\creflabelformat{pluraleq}{#2(#1)#3}
\crefname{section}{Section}{sections}
\crefname{chapter}{Chapter}{chapters}
\crefname{figure}{Figure}{figures}
\crefname{table}{Table}{tables}
\crefname{subfigure}{Figure}{figures}
\creflabelformat{subfigure}{#2#1#3}

\usetikzlibrary{calc}
\usetikzlibrary{arrows}
\usetikzlibrary{positioning}
\usetikzlibrary{intersections}
\usetikzlibrary{decorations.pathreplacing}
\usetikzlibrary{shapes.geometric}

\graphicspath{{figures/}}

\makeatletter
\newcommand*{\dif}{\@ifnextchar^{\DIfF}{\DIfF^{}}}
\def\DIfF^#1{\mathop{\mathrm{\mathstrut d}}\nolimits^{#1}\gobblesp@ce}
\def\gobblesp@ce{\futurelet\diffarg\opsp@ce}
\def\opsp@ce{%
  \let\DiffSpace\!%
  \ifx\diffarg(%
    \let\DiffSpace\relax
  \else
    \ifx\diffarg[%
      \let\DiffSpace\relax
    \else
      \ifx\diffarg\{%
        \let\DiffSpace\relax
      \fi\fi\fi\DiffSpace}
\makeatother

\newcommand*{\dt}[1]{\ensuremath{\frac{\partial #1}{\partial t}}}
\newcommand*{\pd}[2]{\ensuremath{\frac{\partial #1}{\partial{#2}}}}

\newcommand{\set}[2]{\ensuremath{\{\,#1\mid#2\,\}}}

\newcommand*{\vct}[1]{\ensuremath{\boldsymbol{#1}}}

\renewcommand{\div}{\boldsymbol\nabla\cdot}
\newcommand{\grad}{\boldsymbol\nabla}

\newcommand{\sign}{\textup{sign}}

\begin{document}
\pagestyle{empty}
\begin{center}
  {\Large \bf Calculation of interface curvature with the level-set method}
\end{center}
\begin{center}
{\bf Karl Yngve Lervåg} \\[0mm]
{\small Norwegian University of Science and Technology} \\[-1.0mm]
{\small Department of Energy and Process Engineering} \\[-1.0mm]
{\small Kolbjørn Hejes veg 2} \\[-1.0mm]
{\small NO-7491 Trondheim, Norway} \\[-1.0mm]
{\small e--mail: karl.y.lervag@ntnu.no} \\[-1.0mm]
\end{center}
\vspace{5mm}
\parbox{150mm}
{\baselineskip11pt
  {\small
    {\bf Summary} \enspace
    The level-set method is a popular method for interface capturing.  One of
    the advantages of the level-set method is that the curvature and the normal
    vector of the interface can be readily calculated from the level-set
    function.  However, in cases where the level-set method is used to capture
    topological changes, the standard discretization techniques for the
    curvature and the normal vector do not work properly.  This is because they
    are affected by the discontinuities of the signed-distance function
    half-way between two interfaces.  This article addresses the calculation of
    normal vectors and curvatures with the level-set method for such cases.  It
    presents a discretization scheme that is relatively easy to implement in to
    an existing code.  The improved discretization scheme is compared with
    a standard discretization scheme, first for a case with no flow, then for
    a case where two drops collide in a shear flow.  The results show that the
    improved discretization yields more robust calculations in areas where
    topological changes are imminent.
  }
}

\section{Introduction}
The level-set method was introduced by  Osher and Sethian \cite{Osher88}. It is
designed  to implicitly  track moving  interfaces  through an  isocontour of  a
function  defined in  the  entire domain.  In particular,  it  is designed  for
problems in multiple  spatial dimensions in which the topology  of the evolving
interface changes during the course of events, c.f.\ \cite{Sethian03}.

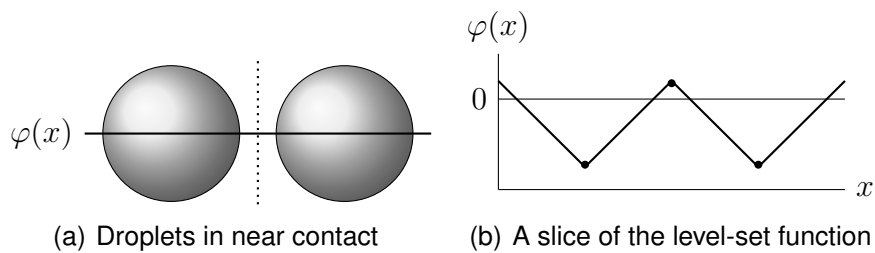
\begin{figure}[bp]
  \centering
  \subfigure[Droplets in near contact]{
    \begin{tikzpicture} 
      [
      scale=0.6,
      drop/.style={shading=ball, ball color=black!10},
      ]

      \shadedraw[drop] (-1.9,0) circle (1.5);
      \shadedraw[drop] ( 1.9,0) circle (1.5);
      \draw[thick] (-3.8,0.0) node[left] {$\varphi(x)$} -- (3.8,0.0);
      \draw[thick, dotted] (0,1.6) -- (0,-1.6);
    \end{tikzpicture}
    \label{fig:1a}}
  \subfigure[A slice of the level-set function]{
    \begin{tikzpicture}[scale=0.6]
      \draw (-3.8,2.5) node[above] {$\varphi(x)$} -- (-3.8,-0.5);
      \draw (-3.8,-0.5) -- ( 3.8,-0.5) node[right] {$x$};
      \draw[thin,black] (-3.8,1.5) node[left] {$0$} -- (3.8,1.5);
      \coordinate (a) at (-3.8,1.9);
      \coordinate (b) at (-1.9,0);
      \coordinate (c) at ( 0.0,1.9);
      \coordinate (d) at ( 1.9,0);
      \coordinate (e) at ( 3.8,1.9);
      \draw[thick] (a) -- (b) -- (c) -- (d) -- (e);

      \fill ( 0.0,1.85) circle (2.5pt);
      \fill (-1.9,0.05) circle (2.5pt);
      \fill ( 1.9,0.05) circle (2.5pt);
    \end{tikzpicture}
    \label{fig:1b}}
  \caption{(a) Two droplets in near contact.  The dotted line marks
  a region where the derivative of the level-set function is not defined.  (b)
  A one-dimensional slice of the level-set function $\varphi(x)$.  The
  dots mark points where the derivative of $\varphi(x)$ is not defined.}
  \label{fig:cdroplets}
\end{figure}

This  article  addresses  the  calculation of  interface  geometries  with  the
level-set method. This method allows us  to calculate the normal vector and the
curvature of an  interface directly as the first and  second derivatives of the
level-set  function.  These  calculations  are  typically  done  with  standard
finite-difference  methods. Since  the level-set  function  is chosen  to be  a
signed-distance function,  in most  cases it  will have areas  where it  is not
smooth. Consider for  instance two colliding droplets where  the interfaces are
captured with the level-set method, see \cref{fig:cdroplets}. The derivative of
the level-set function  will not be defined at the  points outside the droplets
that have  an equal distance  to both droplets. When  the droplets are  in near
contact, this discontinuity  in the derivative will lead  to significant errors
when  calculating  the  interface geometries  with  standard  finite-difference
methods.  For convenience  the  areas  where the  derivative  of the  level-set
function is not defined will hereafter be refered to as kinks.

To the authors  knowledge, this issue was first  described in \cite{Macklin05},
where the  level-set method was  used to model  tumor growth. Here  Macklin and
Lowengrub presented a  direction difference to treat  the discretization across
kinks  for  the normal  vector  and  the  curvature.  They later  presented  an
improved  method where  curve  fitting  was used  to  calculate the  curvatures
\cite{Macklin06}.  This was  further  expanded to  include  the normal  vectors
\cite{Macklin08}.

An  alternative method  to  avoid  the kinks  is  presented in  \cite{Salac08},
where a  level-set extraction  technique is presented.  This technique  uses an
extraction  algorithm  to reconstruct  separate  level-set  functions for  each
distinct body.

Accurate calculation  of the  curvature is important  in many  applications, in
particular  in  curvature-driven  flows.  There are  several  examples  in  the
literature of methods that improve  the accuracy of the curvature calculations,
but that do not consider the  problem with the discretization across the kinks.
The authors in \cite{Wang10} use a coupled level-set and volume-of-fluid method
based on a fixed Eulerian grid, and they use a height function to calculate the
curvatures. In  \cite{Herrmann08} a  refined level-set grid  method is  used to
study two-phase flows on structured and unstructured grids for the flow solver.
An  interface-projected curvature-evaluation  method  is  presented to  achieve
converging calculation of  the curvature. In \cite{Marchandise07}  they adopt a
discontinuous Galerkin  method and a pressure-stabilized  finite-element method
to solve the level-set equation  and the Navier-Stokes equations, respectively.
They  develop a  least-squares  approach  to calculate  the  normal vector  and
the  curvature accurately,  as  opposed to  using a  direct  derivation of  the
level-set function. This method is used in \cite{Desjardins08}, where they show
impressive results for simulation of turbulent atomization.

This article applies  the level-set method to incompressible  two-phase flow in
two dimensions. The direction difference  described in \cite{Macklin05} is used
to calculate  the normal vectors,  and a curvature discretization  is presented
which is based on the  geometry-aware discretization given in \cite{Macklin06}.
The main  advantage of  the present  scheme is  that is  is relatively  easy to
implement, since it requires very little  change to a typical implementation of
the level-set method.

The article starts by briefly  describing the governing equations. It continues
with  a  description  of  the  numerical   methods  that  are  used.  Then  the
discretization schemes for  the normal vector and the  curvature are presented,
followed  by a  detailed description  of the  curvature discretization.  Next a
comparison of  the improved discretization  and the standard  discretization is
made, first on  static interfaces in near contact, then  on two drops colliding
in a shear flow. Finally some concluding remarks are made.

\section{Governing equations}
\subsection{Navier-Stokes equations for two-phase flow}
\begin{figure}[tbp]
  \centering
   \begin{tikzpicture}
     [scale=0.8,border/.style={thick},interface/.style={thick}]
     \draw[border] (0,0) rectangle(10,5);
     \draw[interface] (3,1)
            .. controls (2,1) and (2,4) .. (4,3.5)
            .. controls (6,3) and (5,5) .. (7,4)
            .. controls (9,3) and (9,1.5) .. (8,1.5) node (g1) {} 
            .. controls (4,1.5) and (4,1) .. (3,1);
     \node (g2) [below=0.4cm of g1] {$\Gamma$};
     \draw[->] (g2.west) to [out=150, in=240] (g1.south west);
     \node at (1,1) {$\Omega^+$};
     \node at (4,2) {$\Omega^-$};
   \end{tikzpicture}
  \caption{Illustration of a two-phase domain:  The interface $\Gamma$
  separates the two phases, one in $\Omega^+$ and the other in $\Omega^-$.}
  \label{fig:domain}
\end{figure}
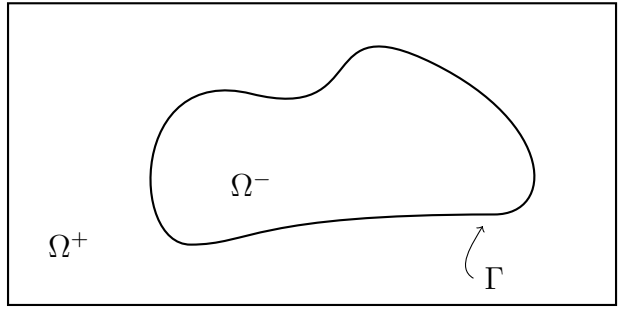

Consider  a two-phase  domain $\Omega=\Omega^+\cup\,\Omega^-$  where $\Omega^+$
and  $\Omega^-$ denote  the  regions  occupied by  the  respective phases.  The
domain is divided by  an interface $\Gamma=\delta\Omega^+\cap\delta\Omega^-$ as
illustrated in  \cref{fig:domain}. The  governing equations  for incompressible
and immiscible two-phase flow in the domain $\Omega$ with an interface force on
the interface $\Gamma$ can be stated as
\begin{align}
  \label{eq:masseq}
  \div\vct u &= 0, \\
  \label{eq:momeq}
  \rho \left( \dt{\vct u} + \vct u\cdot\grad\vct u\right) &= -\grad
  p + \div(\mu\grad\vct u) + \rho\vct f_b + \int_{\Gamma} \sigma\kappa\vct
  n\,\delta(\vct x-\vct x_I(s))\dif s,
\end{align}
where $\vct u$ is the velocity vector, $p$ is the pressure, $\vct f_b$ is the
specific body force, $\sigma$ is the coefficient of surface tension, $\kappa$
is the curvature, $\vct n$ is the normal vector which points to $\Omega^+$,
$\delta$~ is the Dirac Delta function, $\vct x_I$ is a parametrization of the
interface, $\rho$ is the density and $\mu$ is the viscosity.  These equations
are often called the Navier-Stokes equations for incompressible two-phase flow.

It is assumed that the density and the viscosity are constant in each phase,
but they may be discontinuous across the interface.  The interface force and the
discontinuities in the density and the viscosity lead to a set of interface
conditions,
\begin{align}
  [\vct u] &= 0, \\
  [p] &= 2[\mu]\vct n\cdot\grad\vct u\cdot\vct n + \sigma\kappa, \\
  [\mu\grad\vct u] &= [\mu] \big( 
      (\vct n\cdot\grad\vct u\cdot\vct n)\vct n\vct n +
      (\vct n\cdot\grad\vct u\cdot\vct t)\vct n\vct t +
      (\vct n\cdot\grad\vct u\cdot\vct t)\vct t\vct n +
      (\vct t\cdot\grad\vct u\cdot\vct t)\vct t\vct t
    \big), \\
  [\grad p] &= 0,
\end{align}
where $\vct t$ is the tangent vector along the interface and $[\cdot]$ denotes
the jump across an interface, that is
\begin{equation}
  [\mu] \equiv \mu^+ - \mu^-.
\end{equation}
See \cite{Kang00,Hansen05} for more details and a derivation of the interface
conditions.

\subsection{Level-set method}
The interface is captured with the zero level set of the level-set function
$\varphi(\vct x,t)$, which is prescribed as a signed-distance function.  That
is, the interface is given by
\begin{equation}
  \Gamma = \set{(\vct x,t)}{\varphi(\vct x,t)=0}, \quad \vct x\in \Omega,\quad
  t\in \mathbb R^+,
\end{equation}
and for any $t\geq0$,
\begin{equation}
  \varphi(\vct x,t) \left\{
  \begin{array}{ll}
    < 0 & \text{if $\vct x\in\Omega^-$} \\
    = 0 & \text{if $\vct x\in\Gamma$}   \\
    > 0 & \text{if $\vct x\in\Omega^+$}
  \end{array}
  \right..
  \label{eq:level-set-function}
\end{equation}

The interface is updated by solving an advection equation for $\varphi$,
\begin{equation}
  \label{eq:lseq}
  \dt{\varphi} + \vct{\hat u}\cdot\grad\varphi = 0,
\end{equation}
where $\vct{\hat u}$ is the velocity at the interface extended to the entire
domain.  The interface velocity is extended from the interface to the domain by
solving
\begin{equation}
  \label{eq:lsvelext}
  \pd{\vct{\hat u}}{\tau} + S(\varphi)\vct n\cdot\grad\vct{\hat u} = 0,
  \quad \vct{\hat u}_{\tau=0} = \vct u,
\end{equation}
to steady state, c.f.\ \cite{Zhao96}.  Here $\tau$ is pseudo-time and $S$ is
a smeared sign function which is equal to zero at the interface,
\begin{equation}
  S(\varphi) = \frac{\varphi}{\sqrt{\varphi^2+2\Delta x^2}}.
\end{equation}

When \cref{eq:lseq} is solved numerically, the level-set function loses its
signed-distance property due to numerical dissipation.  The level-set function
is therefore reinitialized regularly by solving
\begin{equation}
  \label{eq:lsreinit}
  \begin{split}
    \pd{\varphi}{\tau} + S(\varphi_0)(|\grad\varphi|-1) &= 0, \\
    \varphi(\vct x,0) &= \varphi_0(\vct x),
  \end{split}
\end{equation}
to steady state as proposed in \cite{Sussman94}.  Here $\varphi_0$ is the
level-set function that needs to be reinitialized.

One of the advantages of the level-set method is that normal vectors and 
curvatures can be readily calculated from the level-set function, i.e.
\begin{align}
  \label{eq:norm}
  \vct n &= \frac{\grad\varphi}{|\grad\varphi|}, \\
  \label{eq:curv}
  \kappa &= \div\left(\frac{\grad\varphi}{|\grad\varphi|}\right).
\end{align}

\section{Numerical methods}
The Navier-Stokes equations, \eqref{eq:masseq} and \eqref{eq:momeq}, are solved
by a projection method on a staggered grid as described in
\cite[Chapter~5.1.1]{Hansen05}.  The spatial terms are discretized by the
second-order central difference scheme, except for the convective terms which
are discretized by a fifth-order WENO scheme.  The temporal discretization is
done with explicit strong stability-preserving Runge-Kutta (SSP RK) schemes,
see \cite{Gottlieb01}.  A three-stage third-order SSP-RK method is used for the
Navier-Stokes equations \eqref{eq:masseq} and \eqref{eq:momeq}, and
a four-stage second-order SSP-RK method is used for the level-set equations
\eqref{eq:lseq}, \eqref{eq:lsvelext} and \eqref{eq:lsreinit}.

The method presented in \cite{Adalsteinsson95} is used to improve the
computational speed.  The method is often called the narrow-band method, since
the level-set function is only updated in a narrow band across the interface at
each time step.

The interface conditions are treated in a sharp fashion with the Ghost-Fluid
Method (GFM), which incorporates the discontinuities into the discretization
stencils by altering the stencils close to the interfaces.  For instance, the
GFM requires that a term is added to the stencil on the right-hand side of the
Poisson equation for the pressure.  Consider a one-dimensional case where
$[\rho] = [\mu] = 0$ and where the interface lies between $x_i$ and $x_{i+1}$.
In this case,
\begin{equation}
  \frac{p_{i+1} - 2p_i + p_{i-1}}{\Delta x^2} = f_k \pm
  \frac{\sigma\kappa_{\Gamma}}{\Delta x^2},
  \label{eq:pressure-poisson}
\end{equation}
where $f_k$ is the general right-hand side value and $\kappa_\Gamma$ is the
curvature at the interface.  The sign of the added term depends on the sign of
$\varphi(x_i)$.  See \cite{Kang00} for more details on how the GFM is used for
the Navier-Stokes equations and \cite{Liu00} for a description on how to use
the GFM for a variable-coefficient Poisson equation.

The normal vector and the curvature defined by \cref{eq:norm,eq:curv} are
typically discretized by the second-order central difference scheme, c.f\
\cite{Kang00,Sethian03,Xu06}.  The curvatures are calculated on the grid
nodes and then interpolated with simple linear interpolation to the interface,
e.g.\ for $\kappa_\Gamma$ in \cref{eq:pressure-poisson},
\begin{equation}
  \kappa_\Gamma = \frac{|\varphi_i|\kappa_{i+1}
  + |\varphi_{i+1}|\kappa_i}{|\varphi_i| + |\varphi_{i+1}|}.
  \label{eq:curvature-gamma}
\end{equation}

\begin{figure}[tbp]
  \centering
  \begin{tikzpicture}
    [ 
    axes/.style={thick,gray!150,>=stealth},
    phi/.style={black},
    scale=0.75,
    inner sep=0mm,
    filledcircle/.style={minimum size=3pt,fill=black,circle},
    ]
    
    \node[anchor=south] (posphi) at ( 0.00, 2.0 ) {$\varphi>0$};
    \node[anchor=north] (negphi) at ( 0.00,-2.0 ) {$\varphi<0$};
    \node[anchor=east]  (zerphi) at (-0.25, 0.0 ) {$\varphi=0$};
    \node[anchor=north] (xi)     at ( 5.00,-0.1 ) {$x_i$};
    \node[anchor=west]  (x)      at (10.00, 0.0 ) {$x$};
    \draw[<->,axes] (posphi) -- (negphi);
    \draw[->,axes]  (zerphi) -- (x);
    \foreach \x in {0,...,9}
      \draw (\x,2pt) -- (\x,-2pt);

    \draw[phi] (1.10,-1.45) -- (5.00, 0.40) node[filledcircle] {}
                            -- (8.90,-1.45) ;
  \end{tikzpicture}
  \caption{A level-set function that has one point where the derivative is
  discontinuous.}
  \label{fig:level-set-function}
\end{figure}
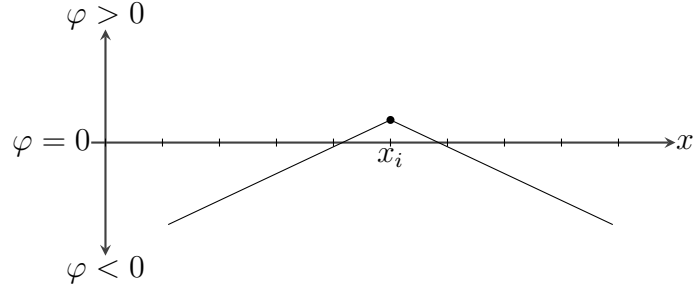

If the level-set method is used to capture non-trivial geometries, the
level-set function will in general contain areas where it is not smooth, i.e.\
kinks.  This is depicted in \cref{fig:level-set-function}, which shows
a level-set function in a one-dimensional domain that captures two interfaces,
one on each side of the grid point $x_i$.  The kink at $x_i$ will lead to
potentially large errors with the standard discretization both for the
curvature and the normal vector.  The errors in the curvature will lead to
erroneous pressure jumps at the interfaces, and the errors in the normal vector
affects both the discretized interface conditions and the advection of the
level-set function.  If the level-set method is used to study for example
coalescence and breakup of drops, these errors may severly affect the
simulations.

It should be noted that the kinks that appear far from any interfaces are
handled by ensuring that the denominators do not become zero, as explained in
\cite[Sections~2.3 to 2.4]{Osher03}.  This works fine, since only the values of
the curvature at the grid nodes adjacent to any interface are used.  Also, the
normal vector only needs to be accurate close to the interface due to the
narrow-band approach.

\section{Improved discretization of geometrical quantities}
The previous section explained why it is necessary to develop new
discretization schemes for the normal vector and the curvature that can handle
kinks in the level-set function.  This section will give a brief presentation
of a better discretization scheme for the normal vector and an overview of an
algorithm to calculate the curvature.

\subsection{The normal vector}
A discretization scheme is presented in \cite{Macklin05} which uses a quality
function to ensure that the differences never cross kinks.  The basic
strategy is to use a combination of central differences and one-sided
differences based on the values of a quality function,
\begin{equation}
  Q(\vct x) = \left| 1 - |\grad\varphi(\vct x)|\right|,
\end{equation}
which is approximated with central differences.  The quality function
effectively detects the areas where the level-set function differs from the
signed-distance function.  Let $Q_{i,j} = Q(\vct x_{i,j})$ and $\eta>0$, then
$Q_{i,j}>\eta$ can be used to detect kinks.  The parameter $\eta$ is tuned such
that the quality function will detect all the kinks.  The value $\eta=0.1$ is
used in the present work.

The quality function is used to define a direction function,
\begin{equation}
  \vct D(\vct x_{i,j}) = (D_x(\vct x_{i,j}), D_y(\vct x_{i,j})),
\end{equation}
where
\begin{equation}
  D_x(\vct x_{i,j}) = \left\{ 
  \begin{array}{ll}
    -1 & \text{if $Q_{i-1,j}<\eta$ and $Q_{i+1,j}\geq\eta$,} \\
     1 & \text{if $Q_{i-1,j}\geq\eta$ and $Q_{i+1,j}<\eta$,} \\
     0 & \text{if $Q_{i-1,j}<\eta$ and $Q_{i,j}<\eta$ and $Q_{i+1,j}<\eta$,} \\
     0 & \text{if $Q_{i-1,j}\geq\eta$ and $Q_{i,j}\geq\eta$ and 
         $Q_{i+1,j}\geq\eta$,} \\
     \text{undetermined} & \text{otherwise.}
  \end{array}
  \right.
  \label{eq:dirfunc}
\end{equation}
$D_y(\vct x_{i,j})$ is defined in a similar manner.  If $D_x$ or $D_y$ is
undetermined, $\vct D(\vct x_{i,j})$ is chosen as the vector normal to
$\grad\varphi(\vct x_{i,j})$.  It is normalized, and the sign is chosen such
that it points in the direction of best quality.  See \cite{Macklin05} for more
details.

The direction difference is now defined as
\begin{equation}
  \partial_x f_{i,j} = \left\{
  \begin{array}{ll}
    \frac{f_{i,j} - f_{i-1,j}}{\Delta x}    & \text{if $D_x(x_i,y_j) = -1$,} \\
    \frac{f_{i+1,j} - f_{i,j}}{\Delta x}    & \text{if $D_x(x_i,y_j) =  1$,} \\
    \frac{f_{i+1,j} - f_{i-1,j}}{2\Delta x} & \text{if $D_x(x_i,y_j) =  0$,}
  \end{array}
  \right.
  \label{eq:Ddifference}
\end{equation}
where $f_{i,j}$ is a piecewise smooth function.  The normal vector is
calculated using the direction difference on $\varphi$, which is equivalent to
using central differences in smooth areas and one-sided differences in areas
close to the kinks.  This method makes sure that the differences do not cross
any kinks, and the normal vector can be accurately calculated even close to
a kink.

\subsection{The curvature}
The curvature is calculated with a discretization that is based on the improved
geometry-aware curvature discretization presented by Macklin and Lowengrub
\cite{Macklin06}.  This is a method where the curvature is calculated at the
interfaces directly with the use of a least-squares curve parametrization of
the interface.  The curve parametrization is used to create a local level-set
function from which the curvature is calculated using standard discretization
techniques.  The local level-set function only depends on one interface and is
therefore free of kinks.

The main difference between the present method and that of Macklin and
Lowengrub is that they calculate the curvature at the interface directly,
whereas the present method instead calculates the curvature at the grid nodes.
In other words, Macklin and Lowengrub calculate $\kappa_\Gamma$ in
\cref{eq:pressure-poisson} directly, whereas the present method calculates
$\kappa_i$ and $\kappa_{i+1}$ with the improved curvature discretization and
then use linear interpolation as described in \cref{eq:curvature-gamma} to find
$\kappa_\Gamma$.  The main motivation behind this difference is that the
present method does not require a significant change to any existing code.
Thus it is relatively straightforward to implement the present method even when
the curvature is needed for more than the Capillary force term in the
Navier-Stokes equations.  An example of such a case is when the curvature is
used to model interfacial flows with surfactant \cite{Xu06}.

An important consequence of the previously explained difference is that it
becomes more important to have an accurate representation of the interface.
The curvature discretization presented here uses monotone cubic Hermite splines
to parametrize the curve.  The least-square parametrization used in
\cite{Macklin06} is only accurate very close to the point where the curvature
needs to be calculated.  The Hermite spline is more accurate along the entire
interface representation.

The algorithm to calculate the curvature at $\vct x_{i,j}$ can be summarized as
follows.  The details are explained in the next section.
\begin{enumerate}
  \item If $Q_{i+n,j+m}\leq\eta$, where $n=-1,0,1$ and $m=-1,0,1$, then it is
    safe to use the standard discretization.  Otherwise continue to the next
    step.
  \item Locate the closest interface, $\Gamma$.
  \item Find a set of points $\vct x_1, \dots, \vct x_n \in \Gamma$.
  \item Create a parametrization $\vct \gamma(s)$ of the points $\vct x_1,
    \dots, \vct x_n$.
  \item Calculate a local level-set function based on the parametrization $\vct
    \gamma(s)$.
  \item Use the standard discretization on the local level-set function to
    calculate the curvature.
\end{enumerate}

\section{Details of the curvature algorithm}
\subsection{Locating the closest interface}
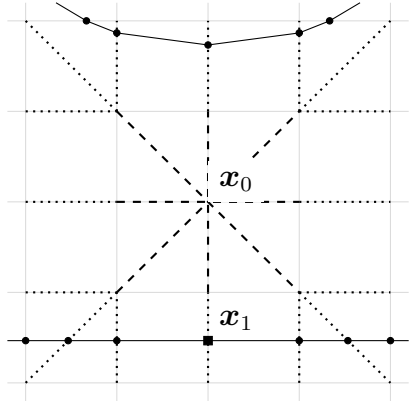
\begin{figure}[tbp]
  \centering
  \begin{tikzpicture}
    [ 
    scale=0.8,
    search1/.style={thick, black, dashed},
    search2/.style={thick, black, dotted},
    grid/.style={very thin,gray!30},
    interface/.style={},
    ]
    
    \draw[grid,step=1.5] (-3.3,-3.3) grid(3.3,3.3);

    \draw[search1] (-1.5, 0.0) -- ( 1.5, 0.0);
    \draw[search1] ( 0.0,-1.5) -- ( 0.0, 1.5);
    \draw[search1] (-1.5,-1.5) -- ( 1.5, 1.5);
    \draw[search1] (-1.5, 1.5) -- ( 1.5,-1.5);
    \draw[search2] (-3.0,-3.0) -- (-1.5,-1.5);
    \draw[search2] (-3.0,-1.5) -- (-1.5,-1.5);
    \draw[search2] (-3.0, 0.0) -- (-1.5, 0.0);
    \draw[search2] (-3.0, 1.5) -- (-1.5, 1.5);
    \draw[search2] (-3.0, 3.0) -- (-1.5, 1.5);
    \draw[search2] (-1.5,-3.0) -- (-1.5,-1.5);
    \draw[search2] ( 0.0,-3.0) -- ( 0.0,-1.5);
    \draw[search2] ( 1.5,-3.0) -- ( 1.5,-1.5);
    \draw[search2] ( 1.5,-1.5) -- ( 3.0,-1.5);
    \draw[search2] ( 1.5,-1.5) -- ( 3.0,-3.0);
    \draw[search2] ( 1.5, 0.0) -- ( 3.0, 0.0);
    \draw[search2] ( 1.5, 1.5) -- ( 3.0, 1.5);
    \draw[search2] ( 1.5, 1.5) -- ( 3.0, 3.0);
    \draw[search2] ( 1.5, 1.5) -- ( 1.5, 3.0);
    \draw[search2] (-1.5, 1.5) -- (-1.5, 3.0);
    \draw[search2] ( 0.0, 1.5) -- ( 0.0, 3.0);

    \draw[interface] (-2.5, 3.3) -- (-2.0, 3.0) -- (-1.5, 2.8) -- ( 0.0, 2.6)
                  -- ( 1.5, 2.8) -- ( 2.0, 3.0) -- ( 2.5, 3.3);
    \draw[interface] (-3.3,-2.3) -- ( 3.3,-2.3);

    \draw[fill] (-2.0, 3.0) circle(1.5pt);
    \draw[fill] ( 2.0, 3.0) circle(1.5pt);
    \draw[fill] (-1.5, 2.8) circle(1.5pt);
    \draw[fill] ( 1.5, 2.8) circle(1.5pt);
    \draw[fill] ( 0.0, 2.6) circle(1.5pt);
    \draw[fill] (-3.0,-2.3) circle(1.5pt);
    \draw[fill] ( 3.0,-2.3) circle(1.5pt);
    \draw[fill] (-2.3,-2.3) circle(1.5pt);
    \draw[fill] ( 2.3,-2.3) circle(1.5pt);
    \draw[fill] (-1.5,-2.3) circle(1.5pt);
    \draw[fill] ( 1.5,-2.3) circle(1.5pt);
    \draw[fill] ( 0.0,-2.3) +(-0.07,-0.07) rectangle +(0.07,0.07);

    \node[above right,fill=white] at (0, 0.0) {$\vct x_0$};
    \node[above right]            at (0,-2.3) {$\vct x_1$};
  \end{tikzpicture}
  \caption{Sketch of a breadth-first search.  The dashed lines depict the edges
  that are searched first, the dotted lines depict the edges that are searched
  next and the solid lines depict two interfaces.  The circular dots mark where
  the algorithm finds interface points, and the rectangular dot marks the point
  which is returned for the depicted case.}
  \label{fig:first-point-2}
\end{figure}

A breadth-first search is used to to identify the closest interface, see
\cref{fig:first-point-2}.  Let $\vct x_0$ denote the starting point and $\vct
x_1$ denote the desired point on the closest interface.  The search iterates
over all the eight edges from $\vct x_0$ to its neighbours and tries to locate
an interface which is identified by a change of sign of $\varphi(\vct x)$.  If
more than one interface is found, $\vct x_1$ is chosen to be the point that is
closest to $\vct x_0$.  If no interfaces are located the search continues at
the next depth.  The search continues in this manner until an interface is
found.  Note that this algorithm does not in general return the point on the
interface which is closest to $\vct x_0$.

The crossing points between the grid edges and the interfaces are found with
linear and bilinear interpolation.  E.g.\ if an interface crosses the edge
between $(i,j)$ and $(i,j+1)$ at $\vct x_I$, the interface point is found by
linear interpolation,
\begin{equation}
  \vct x_I = \vct x_{i,j}+\theta(0,\Delta x),
\end{equation}
where
\begin{equation}
  \theta = \frac{\varphi(\vct x_{i,j})}{\varphi(\vct x_{i,j})-\varphi(\vct
  x_{i,j+1})}.
\end{equation}
In the diagonal cases the interface point is found with bilinear interpolation
along the diagonal.  This leads to
\begin{equation}
  \vct x_I = \vct x_{i,j} + \theta(\Delta x,\Delta x),
\end{equation}
where $\theta$ is the solution of
\begin{equation}
  \alpha_1\theta^2 + \alpha_2\theta + \alpha_3 = 0.
\end{equation}
The $\alpha$ values depend on the grid cell.  For instance, when searching
along the diagonal between $(i,j)$ and $(i+1,j+1)$ the $\alpha$ values will be
\begin{align}
  \alpha_1 &= \varphi_{i,j} 
               - \varphi_{i+1,j} - \varphi_{i,j+1} + \varphi_{i+1,j+1}, \\
  \alpha_2 &= \varphi_{i+1,j} + \varphi_{i,j+1} - 2\varphi_{i,j}, \\
  \alpha_3 &= \varphi_{i,j}.
\end{align}

\subsection{Searching for points on an interface}
When an interface and a corresponding point $\vct x_1$ on the interface are
found, the next step is to find a set of points $\vct x_2,\dots,\vct
x_k,\dots,\vct x_n$ on the same interface.  The points should be ordered such
that when traversing the points from $k=1$ to $k=n$, the phase with
$\varphi(\vct x)<0$ is on the left-hand side.  Note that the ordering of the
points may be done after all the points are found.  Three criteria are used
when searching for new points:
\begin{enumerate}
  \item The points are located on the grid edges.
  \item The distance between $\vct x_k$ and $\vct x_{k+1}$ for all
    $k$ is greater than a given threshold $\mu$.
  \item The $n$ points that are closest to $\vct x_0$ are selected, where $\vct
    x_0=\vct x_{i,j}$ is the initial point where the curvature is to be
    calculated.
\end{enumerate}

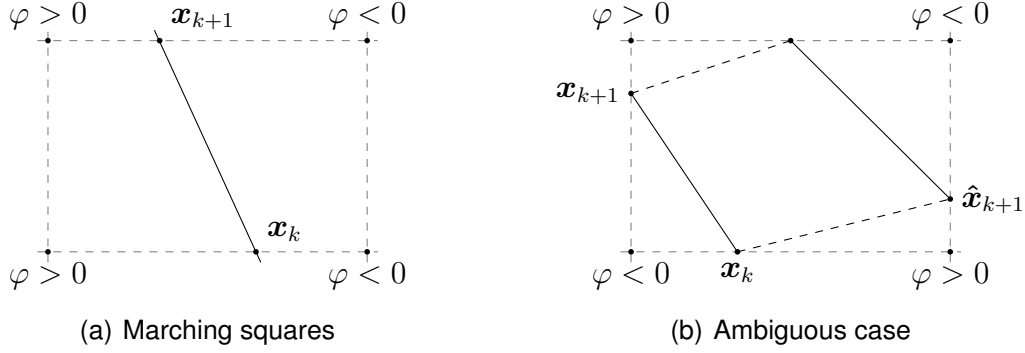
\begin{figure}[tbp]
  \centering
  \subfigure[Marching squares]{
  \begin{tikzpicture}
    [
    scale=0.7,
    rect/.style={gray,thin,dashed} 
    ]

    \draw[rect] (-0.2, 0.0) -- (6.2,0.0);
    \draw[rect] (-0.2, 4.0) -- (6.2,4.0);
    \draw[rect] ( 0.0,-0.2) -- (0.0,4.2);
    \draw[rect] ( 6.0,-0.2) -- (6.0,4.2);
    \fill (0,0) circle(1.5pt) node[below] {$\varphi>0$};
    \fill (6,0) circle(1.5pt) node[below] {$\varphi<0$};
    \fill (0,4) circle(1.5pt) node[above] {$\varphi>0$};
    \fill (6,4) circle(1.5pt) node[above] {$\varphi<0$};
    \draw (2,4.2) -- (4,-0.2) ;
    \fill (2.10,4) circle(1.5pt) node[above right] {$\vct x_{k+1}$};
    \fill (3.91,0) circle(1.5pt) node[above right] {$\vct x_k$};
  \end{tikzpicture}
    \label{fig:marching-square-nor}}
  \hspace{3em}
  \subfigure[Ambiguous case]{
  \begin{tikzpicture}
    [
    scale=0.7,
    krect/.style={gray,thin,dashed} 
    ]

    \draw[krect] (-0.2, 0.0) -- (6.2,0.0);
    \draw[krect] (-0.2, 4.0) -- (6.2,4.0);
    \draw[krect] ( 0.0,-0.2) -- (0.0,4.2);
    \draw[krect] ( 6.0,-0.2) -- (6.0,4.2);
    \fill (0,0) circle(1.5pt) node[below] {$\varphi<0$};
    \fill (6,0) circle(1.5pt) node[below] {$\varphi>0$};
    \fill (0,4) circle(1.5pt) node[above] {$\varphi>0$};
    \fill (6,4) circle(1.5pt) node[above] {$\varphi<0$};

    \draw (2,0) -- (0,3);
    \draw (6,1) -- (3,4);
    \draw[dashed] (2,0) -- (6,1);
    \draw[dashed] (0,3) -- (3,4);
    \fill (2,0) circle(1.5pt) node[below] {$\vct x_k$};
    \fill (6,1) circle(1.5pt) node[right] {$\vct{\hat x}_{k+1}$};
    \fill (0,3) circle(1.5pt) node[left]  {$\vct x_{k+1}$};
    \fill (3,4) circle(1.5pt);
  \end{tikzpicture}
    \label{fig:marching-square-amb}}
  \label{fig:marching-square}
  \caption{(a) The search starts by locating the two points where the interface
  crosses the mesh rectangle.  $\vct x_k$ is the starting point, and if the
  search is counter clockwise it will select $\vct x_{k+1}$ as depicted.  If
  the search is clockwise, it will select $\vct x_{k+1}=\vct x_k$, and the
  search continues in the adjacent mesh rectangle $[x_i,x_{i+1}] \times
  [y_{j-1},y_{j}]$.  (b) An example of an ambiguous case.  The solid black
  lines and the dashed black lines are two equally valid solutions for how the
  interfaces cross the mesh rectangle.  If the search starts at $\vct x_k$ and
  searches counter clockwise, then both $\vct{\hat x}_{k+1}$ and $\vct x_{k+1}$
  are valid solutions.}
\end{figure}

Let $\vct x_k\in\Gamma\cap[x_i,x_{i+1})\times[y_j,y_{j+1})$ be given.  To find
a new point $\vct x_{k+1}$ on $\Gamma$, a variant of the marching-squares
algorithm\footnote{The marching-squares algorithm is an equivalent
two-dimensional formulation of the well known marching-cubes algorithm
presented in \cite{Lorensen87}. The algorithm was mainly developed for use in
computer graphics.} is used.  Given $\vct x_k$ and a search direction which is
either clockwise or counter clockwise, the algorithm searches for all the
points where an interface crosses the edges of the mesh rectangle
$[x_i,x_{i+1}]\times[y_j,y_{j+1}]$.  In most cases there will be two such
points and $\vct x_k$ is one of them.  $\vct x_{k+1}$ is then selected based on
the search direction.  If $\vct x_{k+1}=\vct x_k$, the search is continued
in the adjacent mesh rectangle.  The search process is depicted in
\cref{fig:marching-square-nor}.

In some rare cases the algorithm must handle the ambiguous case depicted in
\cref{fig:marching-square-amb}.  In these cases there are four interface
crossing-points and two solutions.  Either solution is valid, and it is not
possible to say which solution is better.  The current implementation selects
the first solution that it finds, which will be in all practical sense a random
choice.  Note that the ambiguous cases only occur when two interfaces cross
a single grid cell.  The ambiguity comes from the fact that the level-set
method is not able to resolve the interfaces on a sub-cell resolution.

It was found that $n=7$ points where necessary in order to ensure that the
closest points on the interface with respect to the different grid points are
captured with the spline parametrization.

\subsection{Curve fitting}
Cubic Hermite splines are used to fit a curve to the set of points
\begin{equation}
  X_{0,m} = \{\vct x_0, \vct x_1, \dots, \vct x_m\}.
\end{equation}
Let the curve parametrization be denoted $\vct \gamma(s)$ for $0<s<1$.  A cubic
spline is a parametrization where
\begin{equation}
  \vct \gamma(s) = \left\{
  \begin{array}{ll}
    \vct \gamma_1(s) & s_0 \leq s < s_1, \\
    \vct \gamma_2(s) & s_1 \leq s < s_2, \\
    \vdots \\
    \vct \gamma_m(s) & s_{m-1} \leq s \leq s_m,
  \end{array}
  \right.
\end{equation}
where $0=s_0 < s_1 < \dots < s_m = 1$
\begin{equation}
  \vct \gamma(s_i) = \vct x_i,\quad 0\leq i\leq m,
\end{equation}
and each interpolant $\vct \gamma_i(s)=(x_i(s),y_i(s))$ is a third-order
polynomial.  A Hermite spline is a spline where each interpolant is in Hermite
form, see \cite[Chapter~4.5]{Prautzsh02}.  The interpolants are created by
solving the equations
\begin{equation}
  \vct \gamma_i(s) = h_{00}(s)\vct x_{i-1} + h_{01}(s)\vct x_i 
              + h_{10}(s)\vct m_{i-1} + h_{11}(s)\vct m_i,
\end{equation}
for $1\leq i\leq m$, where $\vct m_i$ is the curve tangents and $h_{00},
h_{01}, h_{10}$ and $h_{11}$ are Hermite basis polynomials,
\begin{equation}
  \begin{split}
    h_{00}(s) &= 2s^3 - 3s^2 + 1, \\
    h_{01}(s) &=  s^3 - 2s^2 + s, \\
    h_{10}(s) &=-2s^3 + 3s^2, \\
    h_{11}(s) &=  s^3 -  s^2.
  \end{split}
\end{equation}
The choice of the tangents is non-unique, and there are several possible
options for a cubic Hermite spline.

It is essential that the spline is properly oriented.  This is because we
require to find both the distance and the position of a point on the grid
relative to the spline in order to define a local level-set function.  The
orientation of the spline $\vct \gamma(s)$ is defined such that when $s$
increases, $\Omega^-$ is to the left.

To ensure that our curve is properly oriented, the tangents are chosen as
described in \cite{Fritsch80}.  This will ensure monotonicity for each
component as long as the input data is monotone.  The tangents are modified as
follows.  First the slopes of the secant lines between successive
points are computed,
\begin{equation}
  \vct d_i = \frac{\vct x_{i} - \vct x_{i-1}}{s_i - s_{i-1}}
\end{equation}
for $1\leq i\leq m$.  Next the tangents are initialized as the average of the
secants at every point,
\begin{equation}
  \vct m_{i} = \frac{\vct d_i+\vct d_{i+1}}{2}
\end{equation}
for $1\leq i\leq m-1$.  The curve tangents at the endpoints are set to $\vct
m_0 = \vct d_1$ and $\vct m_m = \vct d_m$.  Finally let $k$ pass from 1 through
$m-1$ and set $\vct m_k=\vct m_{k+1}=0$ where $\vct d_k=0$, and $\vct m_{k}=0$
where $\sign(\vct d_k) \neq \sign(\vct d_{k+1})$.

\subsection{Local level-set function}
The local level-set function, here denoted as $\phi(\vct x_{i,j}) \equiv
\phi_{i,j}$, is calculated at the grid points surrounding and including $\vct
x_0=\vct x_{i,j}$.  The curvature is then calculated with the standard
discretization stencil where $\phi$ is used instead of the global level-set
function, $\varphi$.

A precise definition of $\phi$ is
\begin{equation}
  \phi(\vct x_{i,j}) = \min_s
      \left( \hat d(\vct x_{i,j},\vct \gamma(s)) \right)
  \label{eq:lls}
\end{equation}
where $\hat d (\vct x,\vct \gamma(s))$ is the signed-distance function, which
is negative in phase one and positive in phase two.  This function is
calculated by first finding the minimum distance between $\vct x$ and $\vct
\gamma(s)$ and then deciding the correct sign.  The minimum distance is found
by minimizing the norm
\begin{equation}
  d(\vct x,\vct \gamma(s)) = \|\vct x-\vct \gamma(s)\|_2.
\end{equation}
When $\vct\gamma$ is composed of cubic polynomials as is the case for cubic
Hermite splines, the computation of the distance requires the solution of
several fifth-order polynomial equations.  Sturm's method (see
\cite[Section~11.3]{Waerden03} or \cite[Chapter~XI,§2]{Lang02}) is employed to
locate and bracket the solutions and a combined Newton-Raphson and bisection
method is used to refine them.  The correct sign is found by solving
\begin{equation}
  \sign(\phi(\vct x_{i,j})) = \sign\left( 
  \left(\vct x_{i,j}-\vct \gamma(s)\right) \times
  \vct t_{\vct \gamma}(s) \right)_z,
  \label{eq:sign}
\end{equation}
where $\vct t_{\vct\gamma}(s)$ is the tangent vector of $\vct \gamma(s)$.

\section{Verification and testing}
This section presents results of calculating normal vectors and curvatures with
the improved discretization schemes.  The results are compared with the
standard discretization.  Note that in both the following cases the standard
second-order central differences are used as the standard discretization.

\subsection{A static disc above a rectangle}
\begin{figure}[tbp]
  \centering
  \begin{tikzpicture} 
    [
    scale=1.5,
    hnode/.style = {fill=white,fill opacity=0.5,text opacity=1.0,right=2pt},
    ]
    \draw[fill=gray!20] (0,0.9) circle (0.5);
    \draw[<->] (0,0.9) -- node[above] {$r$} (0.5,0.9);

    \draw[dotted] (0,0.2) parabola (2,1.5);
    \draw[dotted] (0,0.2) parabola (-2,1.5);

    \draw[fill=gray!20] (-2,-1) rectangle (2,0);
    \draw[<->] (0,0.05) -- node[hnode] {$h$} (0,0.35);

    \draw[>=latex',->] (-2,-1) -- node[below] {$x$} (2.5,-1);
    \draw[>=latex',->] (-2,-1) -- node[left] {$y$} (-2,1.5);
  \end{tikzpicture}
  \caption{Initial setup for the circle and line test.  The dotted line depicts
  the kink location.}
  \label{fig:circleandline}
\end{figure}
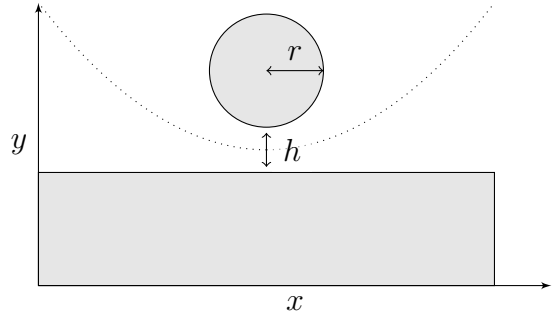

The first case is a simple and static test-case where a disc of radius $r$ is
positioned at a distance $h$ above a rectangle, see \cref{fig:circleandline}.
Only the level-set function and the geometrical quantities are considered.
This means that none of the governing equations are solved
(\cref{eq:masseq,eq:momeq,eq:lseq,eq:lsvelext,eq:lsreinit}).  When $h$ is
small, the kinks along the dotted line will affect the discretization stencils
as has been explained.

The parameters for this case is $r=0.25\ \meter$ and $h = \Delta x$.  The
domain is $1.5\ \meter\times 1.5\meter$, and the straight line is positioned at
$y=0.75\ \meter$.  The grid size is $101\times 101$.

\begin{figure}[tbp]
  \centering
  \includegraphics[width=0.70\textwidth]{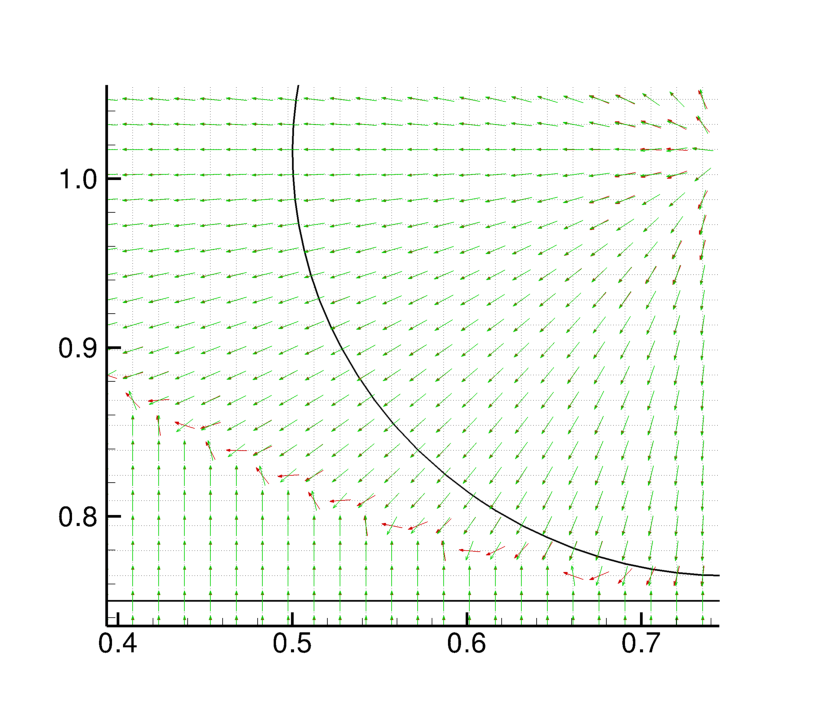}
  \caption{A comparison of the calculated normal vectors between the standard
  discretization in red and the improved method in green.  The thick black
  lines depict the interface.}
  \label{fig:simple-normal-vectors}
\end{figure}

\Cref{fig:simple-normal-vectors} shows a comparison of the calculated normal
vectors.  The standard discretization is depicted with red vectors and the
direction difference is depicted with green vectors.  The figure shows that the
standard discretization yields much less accurate results along the kink region
than the direction difference.

\begin{figure}[tbp]
  \centering
  \subfigure[Standard discretization]{
    \includegraphics[width=0.47\textwidth]{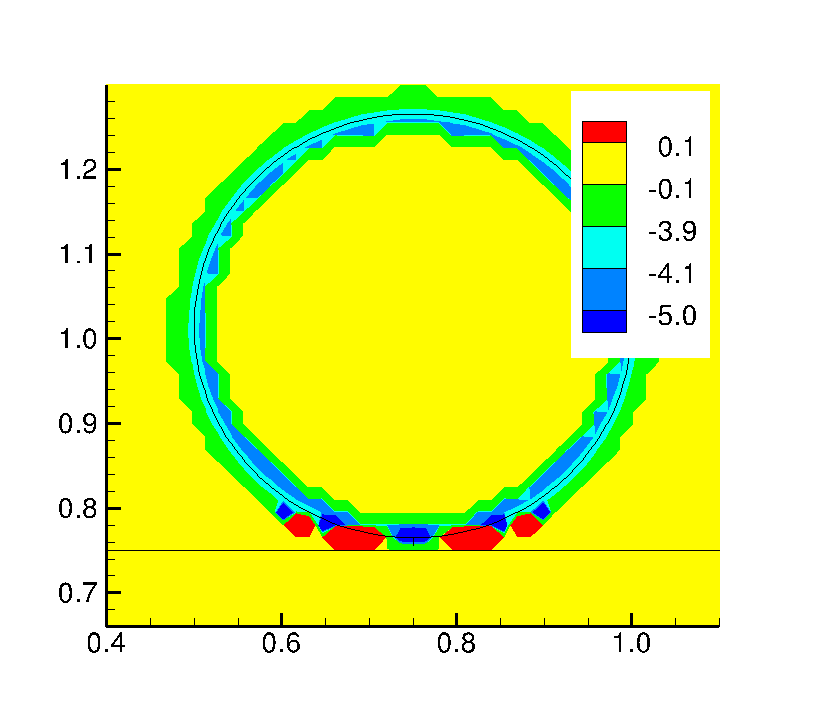}
    \label{fig:before}
    }
  \subfigure[Improved method with curve fitting]{
    \includegraphics[width=0.47\textwidth]{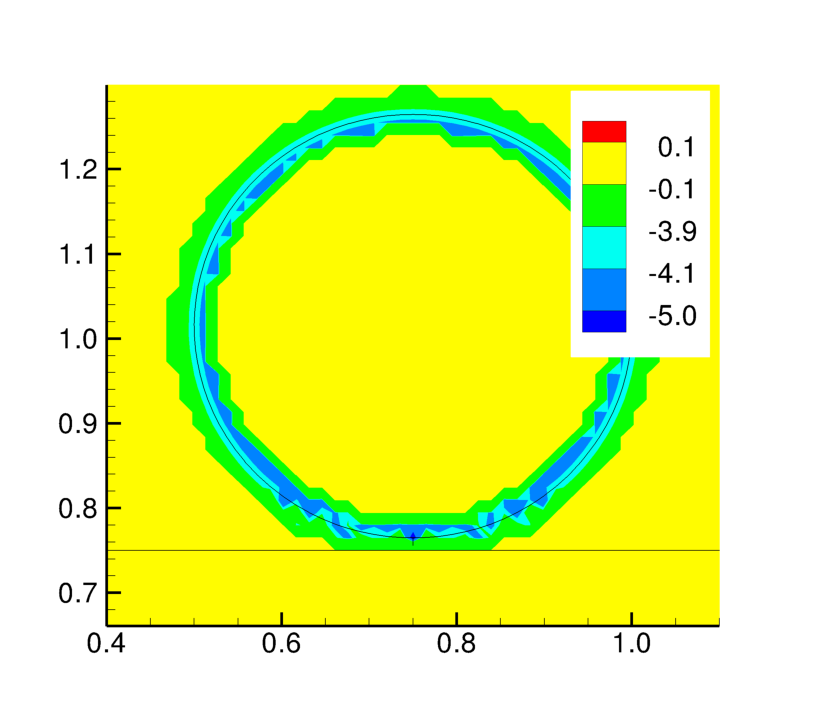}
    \label{fig:after}
    }
  \caption{A comparison of curvature calculations between standard
  discretization and the improved method.  The standard discretization leads to
  large errors in the curvatures in areas that are close to two interfaces.}
  \label{fig:simple-curvatures}
\end{figure}

\Cref{fig:simple-curvatures} shows a comparison of the calculated curvatures.
Note that the curvature is only calculated at grid points adjacent to the
interfaces.  At the grid points where it is not calculated, it is set to zero.
The figure shows that the standard discretization leads to large errors in the
calculated curvatures in the areas that are close to two interfaces.  In
particular note that the sign of the curvature becomes wrong.  The analytic
curvature for this case is $\kappa = -1/r = -4$, and the curvature spikes seen
for the standard discretization is in the order of $|\kappa| \sim
\frac{1}{\Delta x} \simeq 67.3$.  These spikes will lead to large errors in the
pressure jumps through \cref{eq:curvature-gamma}.  The effect of these errors
will become more clear in the next case.

\clearpage
\subsection{Drop collision in shear flow}
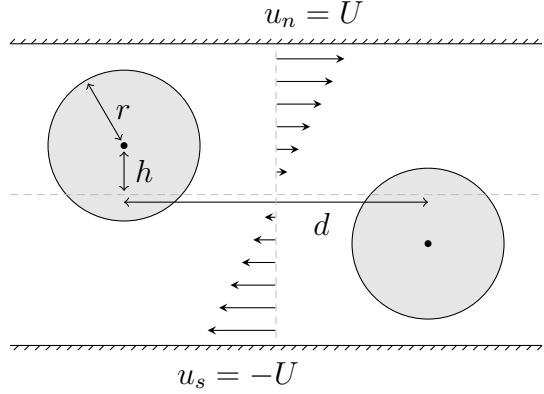
\begin{figure}[tbp]
  \centering
  \begin{tikzpicture}
    [
    wall/.style={
      decoration={border,angle=45,segment length=4},
      postaction={decorate,draw}},
    drop/.style={fill=gray!20},
    axis/.style={very thin, gray!50, densely dashed},
    u/.style={->, >=stealth,},
    ]

    \coordinate (c1) at (-2, 0.65);
    \coordinate (c2) at ( 2,-0.65);
    \draw[drop] (c1) circle(1);
    \draw[drop] (c2) circle(1);
    \fill (c1) circle(1.3pt);
    \fill (c2) circle(1.3pt);
    \draw[<->,rotate=30] (c1) +(0,0.08) -- node[right]{$r$} +(0.0,0.95);
    \draw[<->] (c1) +(0.0,-0.08) -- node[right]{$h$} +(0.0,-0.60);
    \draw[<->] (c1) +(0.0,-0.75) -- ($(c2) + (0.0,0.55)$);
    \node[below] at (0.6,-0.1) {$d$};

    \draw[wall] (-3.5, 2.0) -- ( 3.5, 2.0);
    \node[above=2pt] at ( 0.5, 2.0) {$u_n=U$};
    \draw[wall] ( 3.5,-2.0) -- (-3.5,-2.0);
    \node[below=2pt] at (-0.5,-2.0) {$u_s=-U$};
    \draw[axis] (-3.5, 0.0) -- ( 3.5, 0.0);
    \draw[axis] ( 0.0,-1.9) -- ( 0.0, 1.9);
    \foreach \y/\x in {0.3/0.15,0.6/0.3,0.9/0.45,1.2/0.6,1.5/0.75,1.8/0.9} {
      \draw[u] ( 0.01, \y) -- ( \x, \y);
      \draw[u] (-0.01,-\y) -- (-\x,-\y); }
  \end{tikzpicture}
  \caption{The drop in shear setup.}
  \label{fig:drop-in-shear}
\end{figure}

The second case considers drop collision in shear flow, see
\cref{fig:drop-in-shear}.  The drops both have a radius $r$, and they are
initially placed at a distance $d=5r$ apart in a shear flow where the flow
velocity changes linearly from $u_s=-U<0$ at the bottom wall to $u_n=U$ at the
top wall.  The computational domain is $12r\times8r$, and the grid size is
$241\times 161$.  The size of the grid is chosen to be relatively coarse, such
that the difference between the standard discretization of the curvature and
the curve-fitting based discretization is properly revealed.

The purpose of this case is to study the behaviour of the level-set method, in
particular the calculation of the curvatures, when the drops are in close
proximity.  It is therefore a natural simplification to only consider the case
where the density difference and the viscosity difference of the phases are
zero, i.e.\ there is no jump in density or viscosity across the interface.  

The flow is governed by the Reynolds number and the Capillary number, which in
the current case can be defined by
\begin{align}
  Re &= \frac{\rho Ur}{4\mu}, \\
  Ca &= \frac{\mu U}{4\sigma}.
\end{align}

In the following results the Reynolds and the Capillary numbers were set to
\begin{equation}
  Re = 10,\quad Ca=0.025.
\end{equation}
The choice was made such that the drops would not be severely deformed in the
shear flow.  The radius of the drops was $r = 0.5\ \meter$, and the distance
from the center line to the drop centers was $h = 0.84r = 0.42\ \meter$.

\begin{figure}[tbp]
  \centering
  \begin{tikzpicture}
    \node (nolc1) at (0,0)
      {\includegraphics[width=0.24\textwidth]{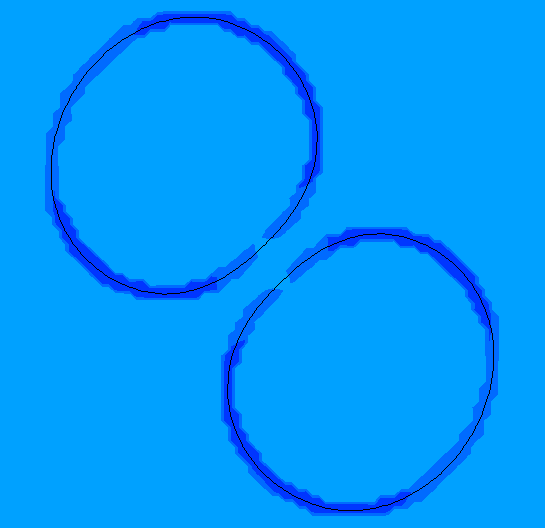}};
    \node (nolc2) [right=-0.15cm of nolc1]
      {\includegraphics[width=0.24\textwidth]{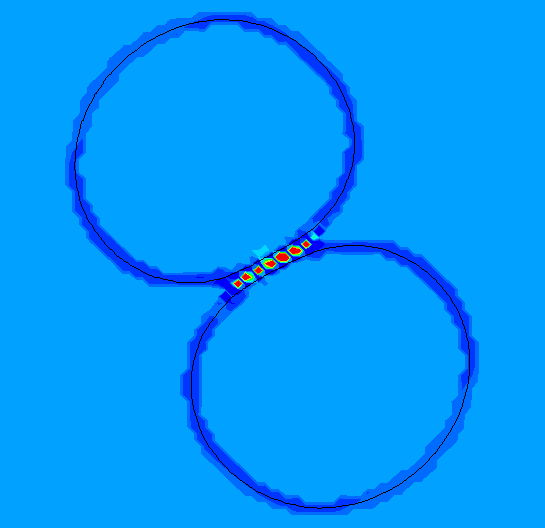}};
    \node (nolc3) [right=-0.15cm of nolc2]
      {\includegraphics[width=0.24\textwidth]{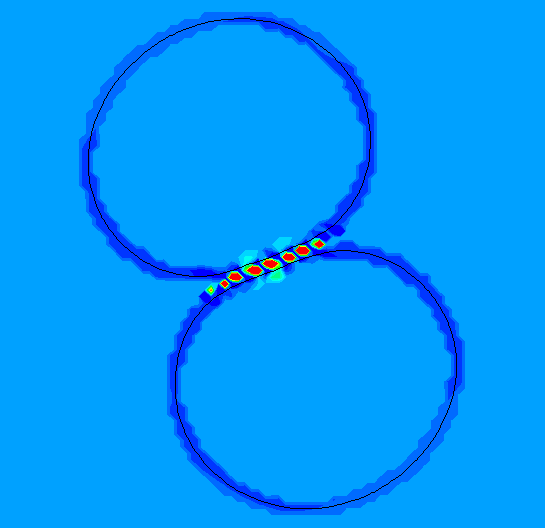}};
    \node (nolc4) [right=-0.15cm of nolc3]
      {\includegraphics[width=0.24\textwidth]{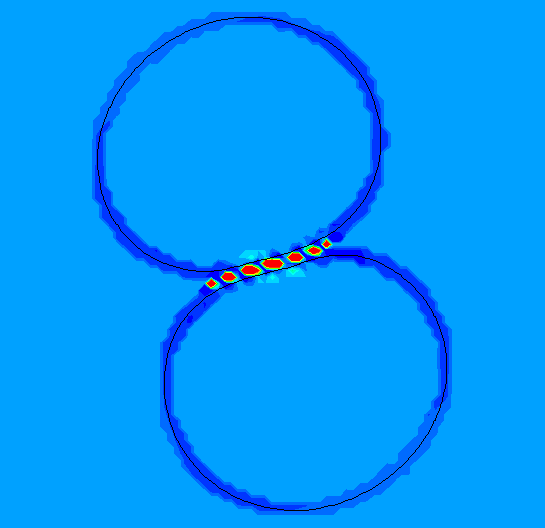}};
    \node (lc1) [below=-0.15cm of nolc1]
      {\includegraphics[width=0.24\textwidth]{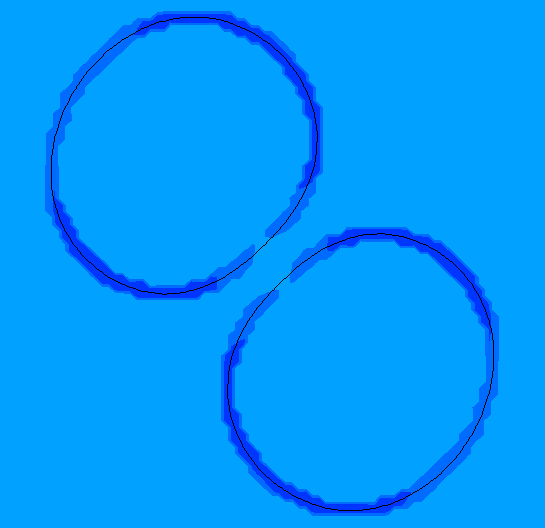}};
    \node (lc2) [right=-0.15cm of lc1]
      {\includegraphics[width=0.24\textwidth]{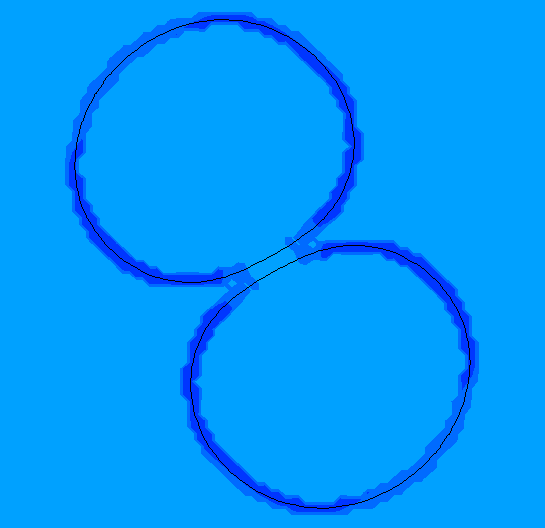}};
    \node (lc3) [right=-0.15cm of lc2]
      {\includegraphics[width=0.24\textwidth]{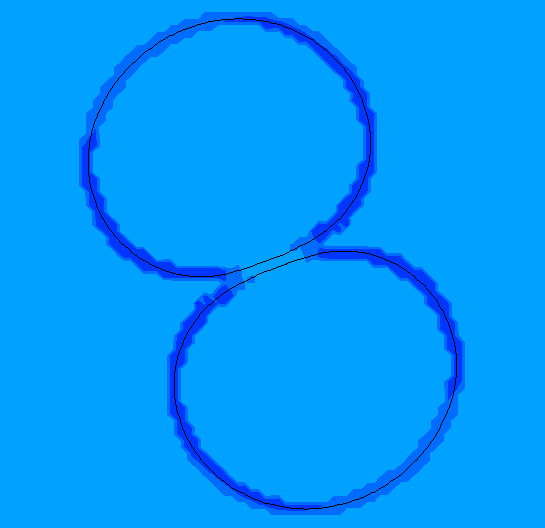}};
    \node (lc4) [right=-0.15cm of lc3]
      {\includegraphics[width=0.24\textwidth]{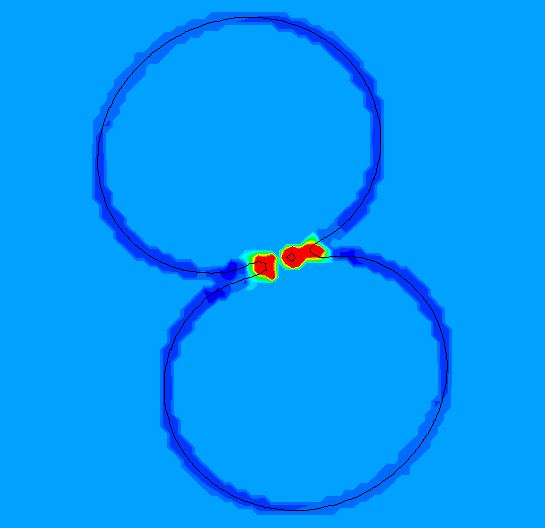}};
    \node [below=-0.13cm of lc1] {$t=2.30\ \second$};
    \node [below=-0.13cm of lc2] {$t=2.60\ \second$};
    \node [below=-0.13cm of lc3] {$t=2.75\ \second$};
    \node [below=-0.13cm of lc4] {$t=2.85\ \second$};

    \node (nolc5) [below=1.0cm of lc1]
      {\includegraphics[width=0.24\textwidth]{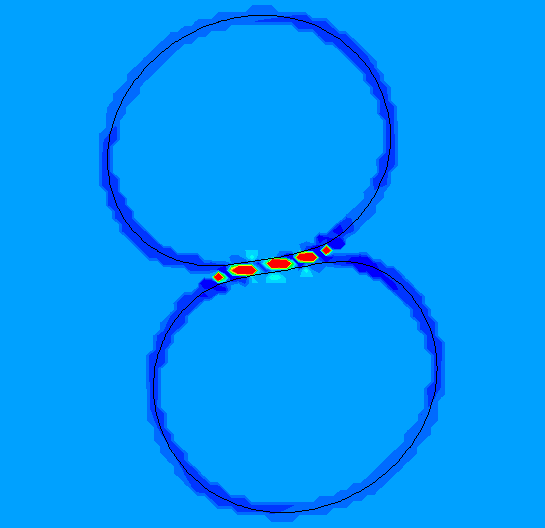}};
    \node (nolc6) [right=-0.15cm of nolc5]
      {\includegraphics[width=0.24\textwidth]{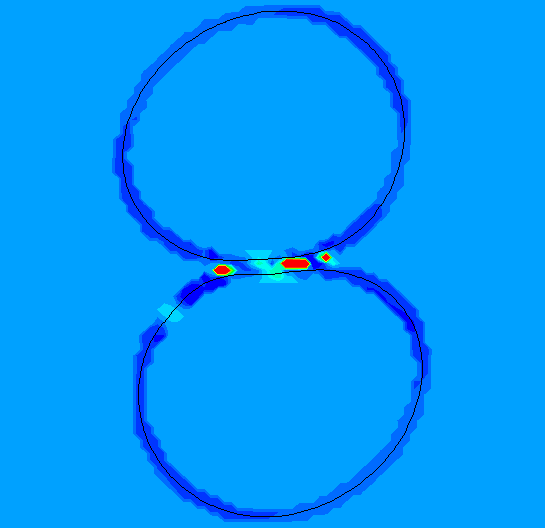}};
    \node (nolc7) [right=-0.15cm of nolc6]
      {\includegraphics[width=0.24\textwidth]{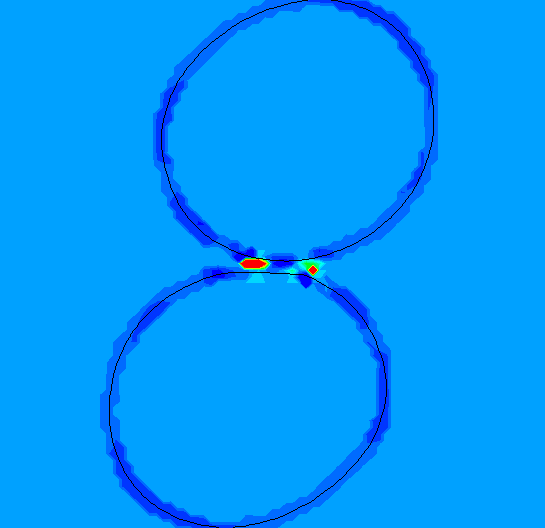}};
    \node (lc5) [below=-0.15cm of nolc5]
      {\includegraphics[width=0.24\textwidth]{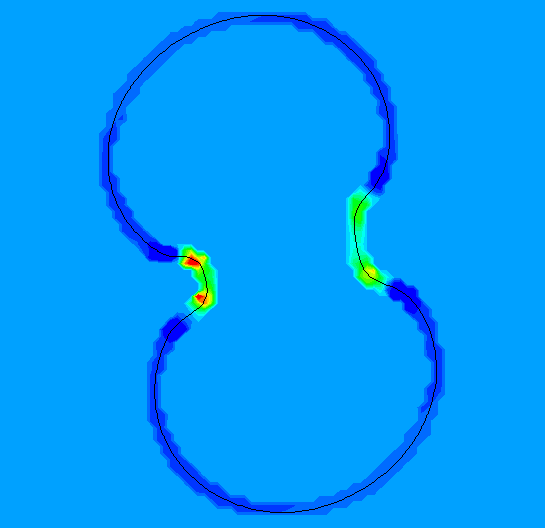}};
    \node (lc6) [right=-0.15cm of lc5]
      {\includegraphics[width=0.24\textwidth]{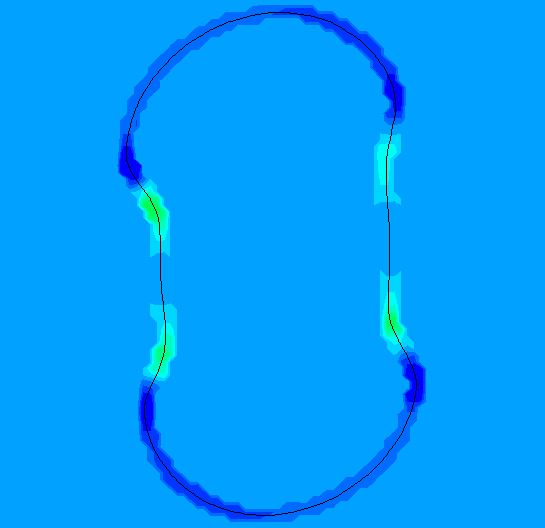}};
    \node (lc7) [right=-0.15cm of lc6]
      {\includegraphics[width=0.24\textwidth]{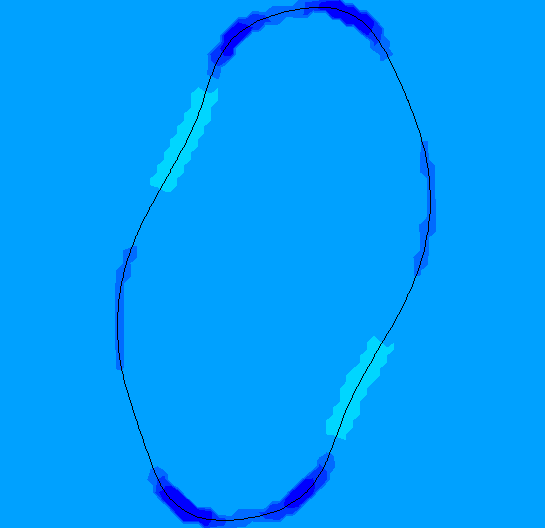}};
    \node [below=-0.13cm of lc5] {$t=2.95\ \second$};
    \node [below=-0.13cm of lc6] {$t=3.10\ \second$};
    \node [below=-0.13cm of lc7] {$t=3.40\ \second$};

    \node (legend) [below=1.3cm of lc4]
      {\includegraphics[width=1.8cm]{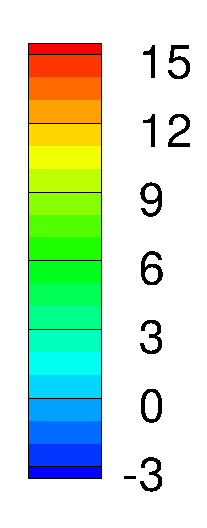}};
    \node [above=-0.3cm of legend] {$\kappa$};

  \end{tikzpicture}
  \caption{A comparison between the standard discretization (top row) and the improved
  discretization (bottom row) of the interface evolution and the curvature
  $\kappa$ of drop collision in shear.}
  \label{fig:ycf84}
\end{figure}
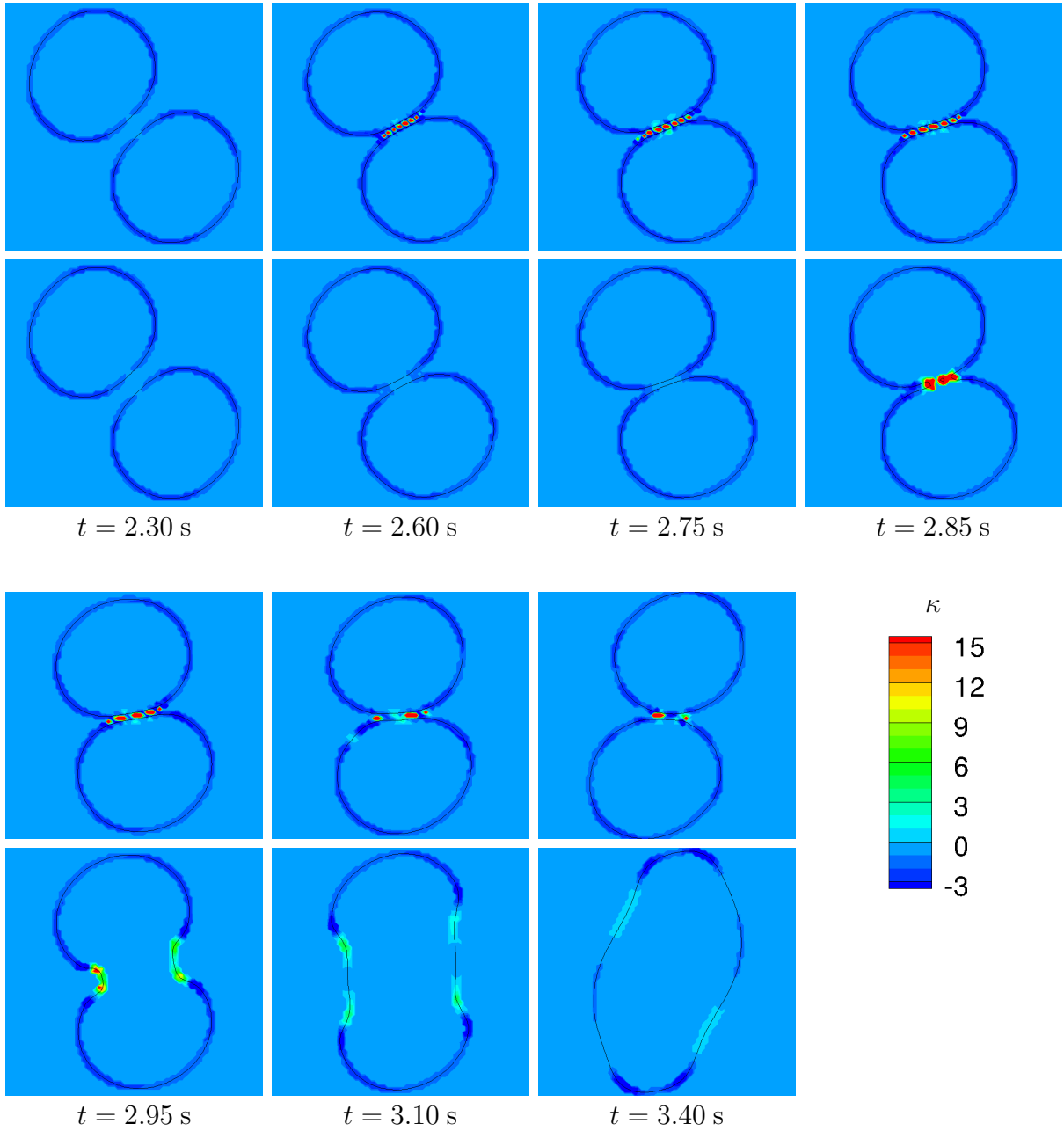

\Cref{fig:ycf84} shows a comparison of the interface evolution and the
curvature between the standard discretization and the improved discretization.
The top and bottom rows show the evolution for the standard discretization and
the improved discretization, respectively.  The kinks between the drops lead to
curvature spikes with the standard discretization, whereas the improved
discretization calculates the curvature along the kink in a much more reliable
manner.  The curvature spikes are seen to prevent coalescence.  This is due to
the effect they have on the pressure field.

\begin{figure}[tbp]
  \centering
  \subfigure[Standard discretization]{
    \includegraphics[width=0.47\textwidth]{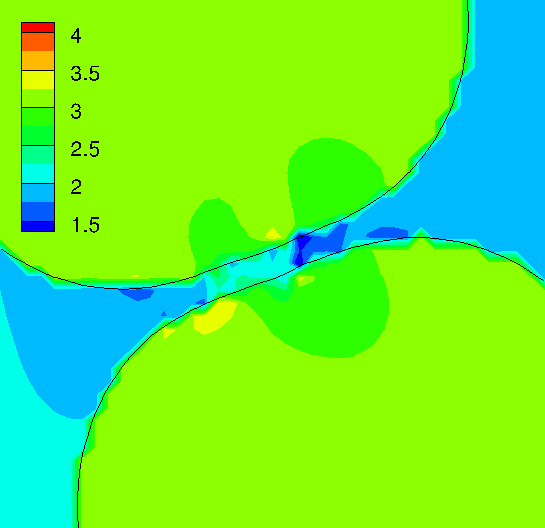}
    \label{fig:ycf84-pressure-nolc}
    }
  \subfigure[Improved method]{
  \includegraphics[width=0.47\textwidth]{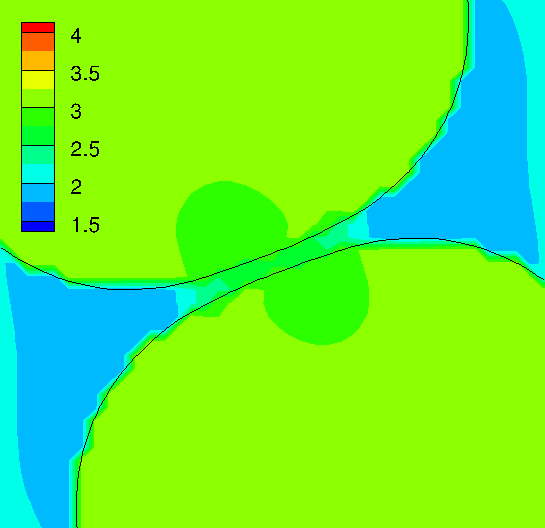}
    \label{fig:ycf84-pressure-lc}
    } \caption{Comparison of the pressure field in the thin film between the
    droplets at $t=2.75\ \second$.  The contour legends indicate the pressure
    in \pascal.}
  \label{fig:ycf84-pressure}
\end{figure}

The errors in the curvature with the standard discretization lead to an
erroneous pressure field between the drops that prevents coalescence, c.f.\
\cref{eq:pressure-poisson}.  \cref{fig:ycf84-pressure} shows the pressure field
at $t=2.75\ \second$.  It can be seen that the pressure field for the standard
discretization is distorted in the thin-film region.  This distortion in the
pressure leads to a flow in the film region which suppresses coalescence.  The
corresponding result for the improved method shows that the pressure is not
distorted.  It is high in the center of the thin-film region and lower at the
edges.  The pressure change induces a flow out of the region which is more as
expected.

\section{Conclusions}
This article has implemented improved discretization schemes for the normal
vector and the curvature of the interface between two phases.  The normal
vector was discretized by the direction difference which is presented in
\cite{Macklin05}.  The curvature was discretized with a scheme that is based on
the geometry-aware discretization presented in \cite{Macklin06}.  The main
advantage of the present discretization method for the curvature is that it is
relatively straightforward to implement in to an existing code since it does
not require a change of the existing framework.

The implementation of the curvature discretization have been described in
detail.  The improved schemes are compared with the standard discretization in
two different cases.  The first case is a direct comparison of the schemes for
a case with no flow.  The second case compares the evolution of two drops
colliding in shear flow.  Both tests demonstrate that the standard
discretization of the normal vector and the curvature leads to erroneous
behaviour at the kink locations.  The second case shows that this behaviour
prevents coalescence from occurring due to an erroneous pressure field.  The
curvature spikes at the kink regions are not observed with the improved
discretization schemes, and coalescence is achieved for the second case.

\section*{Acknowledgements}
This work was financed through the Enabling Low-Emission LNG Systems project,
and the author acknowledge the contributions of GDF SUEZ, Statoil and the
Petromaks programme of the Research Council of Norway (193062/S60).

The author acknowledges Bernhard Müller (NTNU) and Svend Tollak Munkejord
(SINTEF Energy Research) for valuable feedback on the manuscript.  The author
also acknowledges Leif Amund Lie and Eirik Svanes for several good discussions.

\bibliographystyle{mekit09}

\end{document}